\documentstyle[12pt]{article}
\title{  The rare K-decays and $Z^0$-penguin contributions in the 
Topcolor-assisted technicolor models }
\author{ Zhenjun Xiao$^{(1,2,3)}$
\thanks{E-mail: zxiao@ibm320h.phy.pku.edu.cn; and  dphnu@public.zz.ha.cn},
Chongsheng Li$^{(1,2)}$,  Kuangta Chao$^{(1,2)}$ \\  
{\small 1. CCAST(World Laboratory) P.O. Box 8730, Beijing 100080, 
P.R.China} \\
{\small 2. Department of Physics, Peking University,
Beijing, 100871 P.R. China.} \\
{\small 3. Department of Physics, Henan Normal University,
Xinxiang, 453002 P.R. China.} \\  }
\date{February 12, 1999}

\begin{document}
\maketitle
\begin{abstract}
We calculate the new contributions to the rare decays $K^+ \to \pi^+ 
\nu \bar \nu$, $K_L \to \pi^0 \nu \bar \nu$ and  $K_L \to \mu^+ \mu^-$ 
from new $Z^0$-penguin and box diagrams induced by 
the unit-charged scalars $(\tilde{\pi}^\pm, \tilde{H}^\pm, \pi_1^\pm, 
\pi_8^\pm)$  
appeared in the Topcolor-assisted technicolor (TC2) models.  We find that: 
(a) the unit-charged top-pion $\tilde{\pi}^\pm$ and b-pion 
$\tilde{H}^\pm$ can provide large contributions to the rare K-decays 
if they are relatively light;  
(b) the size of mixing elements $D_{L,R}^{ij}$ ($i \neq j$) is strongly 
constrained by the data of $B^0$ meson mixing: 
$|a_R^{ts}|, |a_R^{td}| < 
0.01$ for $a_L^{td}=a_L^{ts}=1/2$ and $m_{\tilde{H}^0}\leq 600GeV$; 
(c) the enhancements to the branching ratios of rare K-decays 
from new scalars can be as large as one order of magnitude; 
(d) there is a strong cancellation between the short- and 
long-distance dispersive part of 
the decay $K_L \to \mu^+ \mu^-$, the constraint on the new short-distance 
part from this decay mode is thus not strong; 
(e) the typical TC2 model under study is generally  consistent with the 
available rare K-decay data. 
\end{abstract}

\vspace{0.5cm}


%\end{document}

\newcommand{\beq}{\begin{eqnarray}}
\newcommand{\eeq}{\end{eqnarray}}

\newcommand{\paa}{\pi_1^{\pm}}
\newcommand{\pbb}{\pi_8^{\pm}}
\newcommand{\pcc}{\tilde{\pi}^{\pm}}
\newcommand{\pcm}{\tilde{\pi}^-}
\newcommand{\pcp}{\tilde{\pi}^+}
\newcommand{\pcz}{\tilde{\pi}^0}
\newcommand{\pdd}{\tilde{H}^{\pm}}

\newcommand{\dsz}{d\overline{s}Z}

\newcommand{\ka}{K^+ \to \pi^+ \nu \overline{\nu}}
\newcommand{\kb}{K_L \to \pi^0 \nu \overline{\nu}}
\newcommand{\kc}{K_L \to \mu^+ \mu^-}
\newcommand{\kcsd}{(K_L \to \mu^+ \mu^-)_{SD}}

\newcommand{\ssa}{\sin^2\theta_W}
\newcommand{\cca}{\cos^2\theta_W}

\newcommand{\xzxt}{X_0(x_t)}
\newcommand{\cztc}{C_0^{New}}
\newcommand{\brka}{B(K^+ \to \pi^+ \nu \bar \nu)}
\newcommand{\brkb}{B(K_L \to \pi^0 \nu \bar \nu)}
\newcommand{\brkc}{B(K_L \to \mu^+  \mu^-)}
\newcommand{\brkcsd}{B(K_L \to \mu^+  \mu^-)_{SD}}
\newcommand{\brkcld}{B(K_L \to \mu^+  \mu^-)_{LD}}

\newcommand{\mpaa}{m_{\pi_1}}
\newcommand{\mpbb}{m_{\pi_8}}
\newcommand{\mpcc}{m_{\tilde{\pi}}}
\newcommand{\mpdd}{m_{\tilde{H}}}
\newcommand{\fpit}{f_{ \tilde{\pi}}}

\newpage

\section{ Introduction}

As is well-known, the study of loop  effects can open an important window 
on electroweak symmetry breaking and physics beyond the standard model(SM). 
The examination of indirect effects of new physics in flavor changing neutral
current (FCNC) processes in rare K- and B-decays 
\cite{buras974,misiak97,buras98,epj981,cpl99} offers a complementary 
approach to the search for direct production of new particles at high 
energy colliders. 

In the SM, the rare K-decays  $\ka$, $\kb$ and $\kc$ 
are all loop-induced semileptonic FCNC processes determined by $Z^0$-penguin 
and W-box diagrams\cite{buras974}.  Since these rare decay modes are 
theoretically clean and highly suppressed in the SM, they may serve as the 
good hunting ground for the new physics beyond the SM. Furthermore, the 
relevant experimental measurements are now reach to a reasonable or even high 
sensitivity\cite{adler97,kek137}, which will help us to test or 
constrain the new physics models through studies about these rare decay 
modes.

In refs.\cite{misiak97,buras98}, the authors studied the FCNC effects on the 
mixing and rare decays for the K and B meson systems in the minimal 
supersymmetric standard model and the two Higgs doublet model. In this paper 
we will investigate the new contributions to the rare K-decays  from the 
$Z^0-$penguin diagrams induced by the unit-charged scalars appeared in 
the TC2 model \cite{hill95}.  

Technicolor (TC) \cite{weinberg76,farhi} is one of the 
important candidates for the mechanism of naturally breaking electroweak 
symmetry. To generate ordinary fermion masses, extended technicolor (ETC)
\cite{susskind79} models have been proposed. In walking technicolor 
theories\cite{holdom81}, the large FCNC problem can be resolved and the 
fermion masses can be increased significantly\cite{holdom81}. The $S$ 
parameter  can be small or even negative in  the walking technicolor 
models\cite{luty93}. To explain the large hierarchy of the quark masses, 
multiscale walking technicolor models (MWTCM) are further constructed
\cite{lane91}. In order to generate   large top quark mass without running 
afoul of either experimental constraints from the $\rho$ parameter and the 
$R_b$ data, the TC2 models was constructed recently
\cite{hill95,lane96,lane98}.

In TC2 model, the relatively light top-pions ($\pcc,\pcz$),  
b-pions ($\tilde{H}^\pm$, $\tilde{H}^0$, 
$\tilde{A}^0$) and other bound states, may provide potentially large 
loop effects in low  energy observables. This is the main 
motivation for us to investigate the contributions to the rare K-decays 
from the penguin and box diagrams induced by the internal exchanges of 
unit-charged top-pions, b-pions and technipions. 

In a previous paper \cite{epj981}, we calculated the $Z^0$-penguin 
contributions to the rare K-decays from technipions $\paa$ and $\pbb$ 
in the MWTCM\cite{lane91} and found that this  
model was strongly disfavored by the relevant data.  

In this paper we calculate the new $Z^0-$penguin contributions to  
the rare K-decays from the top-pions $\pcc$, b-pions $\pdd$ and 
technipions $\paa$ and $\pbb$ appeared in the TC2 model
\cite{hill95}. We firstly evaluate the
$Z^0$-penguin and box diagrams induced by the unit-charged scalars, 
compare the relevant analytical expressions of effective couplings with
the corresponding expressions in the SM, separate the new functions 
$C_0(\pi_i)$ and $C_{NL}{\pi_i}$ ($\pi_i = \pcc, \pdd, \paa, \pbb$) 
which describe the effects of the 
new particles, and finally combine the new functions with 
their counterparts in the SM and use them  directly 
in the calculation for specific decay modes. 

This paper is organized as follows. In Sec.2 we briefly review the basic 
structures of the TC2 models and study the experimental constraints on 
the mixing matrices $D_L$ and $D_R$. In Sec.3 we firstly show the standard 
model predictions for the branching ratios of  rare K-decays, and then 
evaluate the new one-loop Feynman diagrams and extract out the new effective 
$Z^0$-penguin couplings induced by the exchanges of unit-charged scalars. 
In following two sections, we present the numerical results for the 
branching ratios $\brka$, $\brkb$ and $\brkcsd$ with the inclusion of new 
physics effects, and compare the theoretical predictions with the available 
data. The conclusions and discussions are included in the final section.

%%%%%%%
\section{ Basic structure of TC2 models}

Apart from some differences in group structure and/or particle contents,  
all TC2 models \cite{hill95,lane96,lane98} have 
the following common features: (a) strong Topcolor 
interactions, broken near 1 TeV, induce a large 
top condensate and all but a few GeV of the top quark mass, but contribute 
little to electroweak symmetry breaking; (b) TC interactions are 
responsible for electroweak symmetry breaking, and ETC interactions generate 
the hard masses of all quarks and leptons, except that of the top quarks; 
(c) there exist top-pions $\pcc$ and $\pcz$ with a decay constant  
$\fpit \approx 50$ GeV.  In this paper we will chose the most frequently 
studied TC2-I model\cite{hill95} \footnote{In this paper, we use the term 
``TC2-I" model to denote the TC2 model constructed 
by Hill\cite{hill95}.} as the typical TC2 model to estimate 
the contributions to the rare K-decays in question from the relatively light 
unit-charged scalars. It is straightforward to extend the studies in this 
paper to other TC2 models.

\subsection{ TC2-I model, couplings and mass spectrum}

In the TC2-I model\cite{hill95} the dynamics at the scale $\sim$ 1 TeV 
involves the following structure:
\beq
SU(3)_1 \times U(1)_{Y1} \times SU(3)_2 \times U(1)_{Y2} \times SU(2)_L \to 
SU(3)_{QCD} \times U(1)_{EM}\label{group1}
\eeq
where $SU(3)_1 \times U(1)_{Y1}$ ($SU(3)_2 \times U(1)_{Y2}$) generally 
couples preferentially to the third (first and second) generation fermions. 
The breaking (\ref{group1}) typically leaves a  residual global symmetry, 
$SU(3)' \times U(1)'$, implying a degenerate, massive color-octet of 
coloron (ie. the top-gluon) $V_\mu^\alpha$ ($\alpha=1,2,\cdots, 8$)  
and  a color-singlet 
heavy $Z_\mu'$. The gluon $A_\mu^\alpha$ and coloron $V_\mu^\alpha$ ( the SM 
$U(1)_Y$ field $B_\mu$ and the $U(1)'$ field $Z_\mu'$ ), are then defined by 
orthogonal rotations with mixing angle $\theta$ ($\theta'$):
\beq
h_1 \sin \theta &=& h_2 \cos \theta =g_3, \ \ 
 \cot \theta= \frac{h_1}{h_2},  \nonumber\\
q_1 \sin \theta' &=& q_2 \cos \theta' =g_1,  \ \   
\cot \theta'= \frac{q_1}{q_2},  
\eeq
where $h_1$, $h_2$, $q_1$, and $q_2$ are coupling constants of $SU(3)_1$, 
$SU(3)_2$, $U(1)_{Y1}$ and  $U()_{Y2}$, respectively. And $g_3$ ($g_1$) 
is the $SU(3)_C$ ($U(1)_Y$) coupling constant at $\Lambda_{TC}$. In order to 
select the correct top-quark direction for condensation one usually demand 
$\cot \theta >> 1$ and $\cot \theta' >> 1$.   

Both the coloron $V$ and the $Z'$ must be heavier than 1 TeV, according to 
the experimental data from Fermilab Tevatron\cite{burdman97,su97}. After 
integrating out the heavy coloron and $Z'$, the effective four-fermion 
interactions have the form \cite{buchalla96,burdman96} 
\beq
{\cal L}_{eff} =\frac{4 \pi}{M_B^2} \left \{ 
 \left ( \kappa + \frac{2 \kappa_1}{27} \right ) 
 \overline{\psi}_L t_R \overline{t}_R \psi_L 
+  \left ( \kappa - \frac{ \kappa_1}{27} \right )  
\overline{\psi}_L b_R \overline{b}_R \psi_L \right \}, \label{eff1} 
\eeq
where  $\kappa= (g_3^2/4\pi)\cot ^2\theta$ and 
$\kappa_1= (g_1^2/4\pi)\cot ^2\theta'$, and $M_B$ is the mass of  
coloron $V^\alpha$ and $Z'$. 

The effective interactions of (\ref{eff1}) can be written in terms of two
auxiliary scalar doublets $\phi_1$ and $\phi_2$. Their couplings to quarks 
are given by \cite{kominis95}
\beq
{\cal L}_{eff} = \lambda_1 \overline{\psi}_L \phi_1 \overline{t}_R
+ \lambda_2 \overline{\psi}_L \phi_2 \overline{b}_R, \label{eff2} 
\eeq
where $\lambda_1^2 = 4\pi (\kappa + 2\kappa_1/27)$ and 
$\lambda_2^2 = 4\pi (\kappa - \kappa_1/27)$. At energies below the Topcolor 
scale $\Lambda \sim 1$ TeV the auxiliary fields acquire kinetic terms, 
becoming physical degrees of freedom. The properly renormalized  
$\phi_1$ and $\phi_2$ doublets take the form
\beq
\phi_1 = \left ( \begin{array}{cc} 
\fpit + \frac{1}{\sqrt{2}}(h_t + i \pcz) \\ \pcm 
\end{array} \right )
\label{phi1}
\eeq
and 
\beq
\phi_2 = \left ( \begin{array}{cc} 
\tilde{H}^+\\ \frac{1}{\sqrt{2}}(\tilde{H}^0 + i \tilde{A}^0) 
\end{array} \right )
\label{phi2}
\eeq
where $\pcc$ and $\pcz$ are the top-pions, $\tilde{H}^{\pm,0}$ and 
$\tilde{A}^0$ are the b-pions, $h_t$ is the top-Higgs\cite{burdman96}, 
and $\fpit \approx 50 GeV$ is the top-pion decay constant.  

From eq.(\ref{eff2}), the couplings of top-pions 
to t- and b-quark can be written as \cite{hill95}: 
\beq
\frac{m_t^*}{ \fpit } \left[ i\bar{t} t \tilde{\pi}^0 +
  i \overline{t}_R b_L \tilde{\pi}^+
+ \frac{m_b^*}{m_t^*}  \overline{t}_L b_R \tilde{\pi}^+ + h.c.  \right]
\eeq
here, $m_t^* = (1-\epsilon) m_t$ and $m_b^* \approx 1 GeV$ denote the masses 
of top and bottom quarks generated by topcolor interactions. 

For the mass of top-pions, the current $1-\sigma$ lower mass bound from the 
Tevatron data is $\mpcc \geq 150 GeV$\cite{lane96}, while the theoretical 
expectation is $\mpcc \approx (150 - 300 GeV)$\cite{hill95}. 
For the mass of b-pions, 
the current theoretical estimation is $m_{\tilde{H}^0} \approx 
m_{\tilde{A}^0} \approx (100 - 350)  GeV$ and $\mpdd = m_{\tilde{H}^0}^2 
+ 2 m_t^2$ \cite{burdman97}. For the color-singlet $\paa$ and 
color-octet $\pbb$, the current theoretical estimations are $\mpaa 
\geq 50 GeV$ and 
$\mpbb \approx 200GeV$\cite{eichten86,epj981}. In this paper, we  
conservatively consider a little more wider mass ranges of new scalars:
\beq
\mpcc&=& (100 \sim 500) GeV, \ \ \mpdd = (300 - 1000 ) GeV, \ \ 
m_{\tilde{H}^0, \tilde{A}^0} = (150 - 600) GeV, \nonumber\\
\mpaa &=&(50 - 100) GeV, \ \ \mpbb=(100 -300)GeV.
\eeq
For $\fpit$ and $\epsilon$, we use $\fpit=(50 - 60) GeV$ and 
$\epsilon=(0.03 - 0.1)$ \cite{hill95,kominis95}.

The effective Yukawa couplings of ordinary technipions $\paa$ and $\pbb$ to
fermion pairs can be found in refs.\cite{epj981,eichten86,ellis81}.
The relevant gauge couplings of unit-charged scalars to $Z^0$ gauge boson 
are basically model-independent 
and can be written as \cite{epj981,eichten86},
\beq
Z \pi_i^+ \pi_i^-: &&  
-ig\frac{1-2\sin^2\theta_W}{2\cos\theta_W}(p^+ - p^-)\cdot \epsilon',  \\
Z \pi_{8\alpha}^+\pi_{8\beta}^-: &&
 -ig\frac{1-2\sin^2\theta_W}{2\cos\theta_W}(p^+ 
- p^-)\,\delta_{\alpha \beta} \cdot \epsilon',   
\eeq
where $\theta_W$ is the Weinberg angle, $p^+$ and $p^-$ are 
the momenta of relevant scalars, $\epsilon'$ is the polarization vector 
of $Z^0$ gauge boson, and $\pi_i$ denotes the color-singlet scalars 
$\pcc$,  $\pdd$ and $\paa$, respectively.

%%%%%%%%%%%%%%%
\subsection{The square-root ansatz and experimental constraints}

At low energy, potentially large FCNCs arise 
when the quark fields are rotated from their weak eigenbasis to their mass 
eigenbasis, realized by the matrices $U_{L,R}$ for the up-type quarks, 
and by $D_{L,R}$ for the down-type quarks. When we make the replacements, 
for example, 
\beq
b_L \to  D_L^{bd} d_L  +   D_L^{bs} s_L + D_L^{bb} b_L, \\
b_R \to D_R^{bd} d_R + D_R^{bs} s_R + D_R^{bb} b_R,
\eeq
the FCNC interactions will be induced. In TC2-I model, the corresponding  
flavor changing effective Yukawa  couplings  are 
\beq
\frac{m_t^*}{\fpit} \left[ 
i\tilde{\pi}^+ ( D_L^{bs}\bar{t}_R  s_L +  D_L^{bd}\bar{t}_R d_L) + 
i\tilde{H}^+ ( D_R^{bs} \bar{t}_L s_R +   D_R^{bd}\bar{t}_L d_R) 
+ h.c. \right ].
\eeq

Although there are many discussions about the mixing matrixes
in the TC2 models\cite{hill95,buchalla96,kominis95,lane96b}, 
there exist no  ``standard" mixing matrixes currently.  
In the literature, authors usually 
use the ``square-root ansatz":  to take the square root of the standard 
model CKM matrix ($V_{CKM}=U_L^+ D_L$) 
as an indication of the size of realistic mixings. It should be denoted that 
the square root ansatz must be modified because of the 
strong constraints from the data of $B^0 - \overline{B^0}$ mixing
\cite{kominis95,lane96b}.

In the SM, the $B^0$ meson mixings are occurred in second order weak 
interactions. There is evidence for $B^0-\overline{B^0}$ mixing with
$\Delta M_{B_d}= (3.05 \pm 0.12)\times 10^{-10} MeV$\cite{pdg98}, and for 
$B_s^0-\overline{B_s^0}$ mixing, with $\Delta M_{B_s} 
> 6 \times 10^{-9} MeV$ ($CL=95\%$)\cite{pdg98}. 

In TC2 models, the neutral 
scalars $\tilde{H}^0$ and $\tilde{A}^0$ can induce a contribution to the 
$B_q^0-\overline{B_q^0}$ ($q=d, s$) mass difference
\cite{buchalla96,kominis95}
\beq
\frac{\Delta M_{B_q}}{M_{B_q}} 
= \frac{7}{12}\frac{m_t^2}{\fpit^2 m^2_{\tilde{H}^0}}
\delta_{bq}B_{B_q} F_{B_q}^2
\label{deltabd}
\eeq
where $M_{B_q}$ is the mass of $B_q$ meson, $F_{B_q}$ is the $B_q$-meson 
decay constant, $B_{B_q}$ is the renormalization group invariant parameter, 
and $\delta_{bq} \approx |D_L^{bq}D_R^{bq}|$. 
For $B_d$ meson, using the data of $\Delta M_{B_d}=(3.05 \pm 0.12)\times 
10^{-10} MeV$\cite{pdg98} and setting $\fpit =50GeV$, 
$\sqrt{B_{B_d}}F_{B_d}=200 MeV$\cite{buras974},  one has the bound
\beq
\delta_{bd} \leq 0.76 \times 10^{-7} 
\label{boundbd}
\eeq
for $m_{\tilde{H}^0} \leq 600 GeV $. 
This is an important and strong bound on the product of mixing elements 
$D_{L,R}^{bd}$. As pointed in ref.\cite{buchalla96}, 
if one naively uses the square-root ansatz for {\em both}  $D_L$ and $D_R$, 
the bound (\ref{boundbd}) is violated by about 2 orders of magnitudes. 
As shown in ref.\cite{buchalla96}, the ``triangular texture" of the mixing 
matrix may provide a natural suppression of the effect 
by producing approximately diagonal $D_L$ {\em or} $D_R$ matrices. This 
will give $\delta_{bd} \approx 0$ and avoids the bound.

Numerically, if we use the square-root ansatz for $D_L$ itself and 
assume that $D_{L,R}^{bd}/V_{td} =a_{L,R}^{td}$, then the bound 
(\ref{boundbd}) can be written in a new form 
\beq
|a_R^{td}| \leq 1.7 \times 10^{-3}, \label{arl1}
\eeq
for $a_L^{td}=0.5$ and $m_{\tilde{H}^0} \leq 600 GeV$. It is obviously a 
very strong constraint on $D_R^{bd}$.

For $B_s$ meson, the available date is only a lower bound on $\Delta 
m_{B_s}$ \cite{pdg98}: 
\beq
\Delta M_{B_s} > 6 \times 10^{-9} MeV =19.6 \Delta M_{B_d}. \label{ratio1}
\eeq
But one can get a reliable estimation of $\Delta M_{B_s}$ from
its relation with $\Delta M_{B_d}$. In the ratio of $B_s$ and $B_d$ 
mass differences, many common factors cancel, and we have\cite{pdg98}
\beq
\Delta M_{B_s} = \Delta M_{B_d} \frac{M_{B_s}}{M_{B_d}}
\frac{B_{B_s}F_{B_s}^2}{B_{B_d}F_{B_d}^2} 
\frac{|V_{tb}^*\cdot V_{ts}|^2}{|V_{tb}^*\cdot V_{td}|^2}
= 24.9 \Delta M_{B_d}, \label{ratio2}
\eeq
where we have used $M_{B_d}=5.279GeV$, 
$M_{B_s}=5.369GeV$, and $B_{B_s}/B_{B_d}=1.01 \pm 0.04$ and 
$F_{B_s}/F_{B_d}=1.15 \pm 0.05$ from lattice QCD \cite{flynn96}. 
Using eqs.(\ref{deltabd},\ref{ratio2}) and assuming $D_{L,R}^{bs}/V_{ts} =a_{L,R}^{ts}$, 
we have 
\beq
|a_R^{ts}| \leq 1.6 \times 10^{-3}, \label{arl3}
\eeq
for $a_L^{ts}=0.5$ and $m_{\tilde{H}^0} \leq 600 GeV$. 
From eqs.(\ref{ratio1}) and (\ref{ratio2}) we believe that the true 
value of $\Delta M_{B_s}$ should be within the range of 
$(19.6 - 24.9) \Delta 
M_{B_d}$. The inclusion of uncertainties of the data and input 
parameters will weaken the above constraints, but can not change them 
greatly. It is thus reasonable to expect that both $|a_R^{td}|$ and 
$|a_R^{ts}|$ can not be larger than $0.01$ for $a_L^{tj}\approx 1/2$ 
($j=d, s$) and $m_{\tilde{H}^0} \leq 600 GeV$. We conservatively use $0.01$  
as the upper bound on both $a_R^{ts}$ and $a_R^{td}$ afterwards.

In this paper we  assume that all elements of $D_L$ and $D_R$ are real 
because we do not study the CP violation here. We will consider the 
following two typical cases in numerical calculation:
\begin{itemize}
\item
Case-A: Assuming $\epsilon=0.05 $ and $\fpit=50 GeV$, 
using the square-root ansatz for $D_L$: $ a_L^{ts}= a_L^{td}= 1/2$,  but 
setting $a_R^{ts}=a_R^{td}=0.01$.

\item
Case-B:  Assuming $\epsilon=0.05$ and $\fpit=50 GeV$, 
and using the square-root ansatz for {\em both} $ D_L$ and $D_R$: 
$a_L^{tj}=1/2,  |a_R^{tj}| = 1/2$ ($j=d, s$).
\end{itemize}

The Case-A is consistent with the constraints from the data of $B_q^0 - 
\overline{B_q^0}$ ( $q=d,s$) mixing. For Case-B, it violates the constraints 
(\ref{arl1},\ref{arl3}).  
We still consider the Case-B here in order to see  what will happen 
if we use the popular square root ansatz for both $D_L$ and $D_R$.

In this paper, we  fix the following relevant parameters 
\cite{buras974,pdg98} and use them as the standard input(SIP):
\beq
M_W&=&80.41 GeV,\;  \alpha_{em}=1/129,\; \ssa =0.23,\;  
m_c \equiv  \overline{m_c}(m_c)=1.35GeV, \nonumber\\ 
m_t &\equiv& \overline{m_t}(m_t)=170GeV, \; \mu_c=1.3GeV,\; \mu_t =170GeV, 
\nonumber \\
\Lambda^{(4)}_{\overline{MS}}&=&0.325GeV, \; 
\Lambda^{(5)}_{\overline{MS}}=0.225GeV, \; 
A=0.84, \;  \lambda=0.22, \;  \rho=0,\;  \eta=0.36
\label{sip}
\eeq
where the $A, \lambda, \rho$ and $\eta$ are Wolfenstein parameters at 
the leading order.  For $\alpha_s(\mu)$ we use the two-loop expression as 
given in second paper of ref.\cite{buras974}.

%%%%%%%%%
%%%%%%%%%
\section{ Rare K-decays and new physics effects}

In this paper we use the ``Penguin Box expansion'' (PBE) approach
\cite{buras91}. One important advantage of this PBE approach is that 
the rare K-decays in question depend only on one or two basic, universal, 
process independent functions. At next-to-leading order(NLO), such functions 
are $X(x_t)$ and $X_{NL}^l$ for decays $\ka$ and $\kb$, and $Y(x_t)$ and 
$Y_{NL}$ for the short-distance part of the decay $\kc$. 
The functions $X(x_t)$ and $Y(x_t)$ describe 
the dominant $m_t$-dependent contributions, while the functions $X_{NL}^l$ 
and $Y_{NL}$ describe the $m_t$-independent contributions stemming from 
internal quarks other than the top quark (usually known as the charm part).

\subsection{ Rare K-decays in the SM}

In the SM, the rare K-decays $\ka$, $\kb$ and the short-distance part of 
$\kc$ have been studied in great detail and was summarized for instance in a 
new review paper \cite{buras974}. At the leading order(LO), the contributions 
to the rare K-decays from 
$Z^0$-penguin and W-box diagrams are controlled by the functions $C_0(x_i)$
and $B_0(x_i)$ ($i = u, c,t$), which were evaluated long time ago by 
Inami and Lim\cite{inami81}. In recent years, the NLO
corrections are calculated systematically by many authors\cite{buras974}. 
The great progress 
in both the theoretical and experimental investigations enable us now to 
study the new physics effects on the rare K-decays. 

At the NLO level, the effective Hamiltonian for $\ka$, $\kb$ and $\kcsd$  
can be written as \cite{buras974}, 
\beq
{\cal H}_{eff}(\ka) = \frac{\alpha_{em} G_F}{2\sqrt{2}\pi \ssa}
  \sum_{l=e, \mu, \tau} \left [  \lambda_t X(x_t) + \lambda_c
 X^l_{NL} \right ] 
(\overline{s}d)_{V-A}(\overline{\nu_l}\nu_l)_{V-A},  \label{effa} \\
{\cal H}_{eff}(\kb) = \frac{\alpha_{em} G_F}{2\sqrt{2}\pi \ssa}
\lambda_t X(x_t) (\bar s d)_{V-A}(\bar\nu\nu)_{V-A} + h.c.,  
\label{effb}\\
{\cal H}_{eff}(\kcsd) = - \frac{\alpha_{em}G_F}{2\sqrt{2}\pi \ssa}
  \left[  \lambda_t Y(x_t) + \lambda_c Y_{NL} \right] 
(\overline{s}d)_{V-A}(\overline{\mu}\mu)_{V-A} + h.c., 
\label{effc}
\eeq
where $\alpha_{em}$ is the electromagnetic fine-structure constant, 
$G_F=1.16639\times 10^{-5} GeV^{-2}$ is the Fermi coupling constant, 
and $\lambda_i=V^*_{is}V_{id}$ and 
$V_{ij}$ ($i=u, c, t; j=d,s,b$) are the elements of the CKM mixing matrix 
in the SM. 
The functions $X(x_t)$, $Y(x_t)$, $X_{NL}^l$ and $Y_{NL}$ are 
\beq
X(x_t) &=& 
C_0(x_t)-4B_0(x_t) + \frac{\alpha_s}{4\pi} X_1(x_t), \label{xxt} \\
Y(x_t) &=& 
C_0(x_t)-B_0(x_t) + \frac{\alpha_s}{4\pi} Y_1(x_t), \label{yxt} \\
X_{NL}^l &=& C_{NL} -4 B_{NL}^{1/2}, \label{xnl}  \\ 
Y_{NL} &=& C_{NL} - B_{NL}^{-1/2}\label{ynl}
\eeq
where the functions $C_0(x_t)$ and $B_0(x_t)$ are leading top quark 
contributions through  the $Z^0$-penguin and W-box diagrams respectively, 
$X_1(x_t)$ and $Y_1(x_t)$ are NLO QCD corrections, $C_{NL}$ is the 
$Z^0$-penguin part in charm sector, and finally  the functions 
$B_{NL}^{1/2}$ and $B_{NL}^{-1/2}$ are the W-box contributions in the charm 
sector, relevant for the case of final state neutrinos (leptons) with weak 
isospin $T_3=1/2$ ($-1/2$) respectively. 

Using the effective Hamiltonians $(\ref{effa},\ref{effb})$ and summing over 
the three neutrino flavors we have \cite{buras974}
\beq
B(\ka) &=& \kappa_+ \cdot \left[ 
\left( \frac{Im \lambda_t}{\lambda^5} X(x_t) \right)^2 
+\left( \frac{Re \lambda_c}{\lambda}P_0(X) + 
\frac{Re \lambda_t}{\lambda^5}X(x_t) \right)^2 \right], \label{bra}\\
\brkb &=& \kappa_L \cdot \left( \frac{Im \lambda_t}{\lambda^5} X(x_t) 
\right)^2, \label{brb} 
\eeq
where $\kappa_+ = 4.11 \times 10^{-11}$ and $\kappa_L =1.80 \times 10^{-10}$
as given in ref.\cite{buras974}.

For the short-distance part of $\kc$, we have 
\beq
\brkcsd = \kappa_{\mu} \cdot \left[ 
\frac{Re \lambda_c}{\lambda}P_0(Y) + 
\frac{Re \lambda_t}{\lambda^5}Y(x_t) \right]^2 \label{brcsd}
\eeq
where $\kappa_{\mu} =1.68 \times 10^{-9} $\cite{buras974}.
The functions $P_0(X)$ and $P_0(Y)$ in eqs.(\ref{bra},\ref{brcsd}) 
describe the contributions from the charm sector and have been defined 
in ref.\cite{buras974}, 
\beq
P_0(X) &=& \frac{1}{\lambda^4} \left[ \frac{2}{3}X_{NL}^e + 
\frac{1}{3} X_{NL}^{\tau} \right ] \\
P_0(Y)&=&\frac{Y_{NL}}{\lambda^4}. 
\eeq

The explicit expressions for the functions 
$C_0(x_t)$, $B_0(x_t)$, $X_1(x_t)$, $Y_1(x_t)$, 
$C_{NL}$, $B_{NL}^{1/2}$ and $B_{NL}^{-1/2}$ can be found for instance 
in refs.\cite{buras974}. For the convenience of the reader, we present
these functions in Appendix A.

%%%%%%%%%%%%%%% 
\subsection{New $Z^0$-penguin contributions in the TC2-I model }
 
For the rare K-decays under consideration, the new physics will manifest 
itself by modifying the functions $X(x_t)$ and $Y(x_t)$, 
as well as the functions $X_{NL}$ and $Y_{NL}$ in 
effective Hamiltonians (\ref{effa},\ref{effb},\ref{effc}). 

The new one-loop diagrams can be obtained from the diagrams in the SM by 
replacing the internal $W^{\pm}$ lines with the unit-charged scalar 
lines, as shown in Fig.1. 
The color-octet $\pbb$ does not couple to the $l\nu$ lepton pairs, 
and therefore does not present in the box diagrams. For the color-singlet 
scalars, they do couple to $l\nu$ pairs through box diagrams, but the 
relevant couplings are strongly suppressed  by the lightness of $m_l$.
Consequently, we can safely neglect  the tiny contributions from 
those scalars through the box diagrams. 

In ref.\cite{epj981}, we evaluated the new one-loop 
$Z^0-$penguin and W-box diagrams for the induced $\dsz$ couplings due to
the exchange of unit-charged technipions $\paa$ and $\pbb$ in the 
multiscale walking technicolor model\cite{lane91}. In this paper we use the 
same method and follow the same procedure to evaluate the one-loop 
diagrams induced by top-pions $\pcc$ and b-pions $\pdd$. 

We will use dimensional regularization to regulate 
all the ultraviolet divergences in the virtual loop corrections 
and adopt the $\overline{MS}$ renormalization scheme.  It is easy to show 
that all ultraviolet divergences are canceled for scalars $\pcc$, $\pdd$, 
$\paa$ and $\pbb$ respectively,  and therefore the total sum is finite.

By analytical evaluations of the Feynman diagrams, we find the effective 
$\dsz$ vertex induced by the charged top-pion exchange,
\beq
\Gamma^{I}_{Z_{\mu}} = 
\frac{1}{16 \pi^2}\frac{g^3}{\cos\theta_W}\; \sum_{j} \lambda_j\,
\overline{s_L}\, \gamma_{\mu}\, d_L\; C_0(\xi_j), \label{bszc}  
\eeq
with
\beq
C_0(\xi_j)&=& \frac{a_L^{ts*}a_L^{td}\; \mpcc^2 }{2\sqrt{2}\fpit^2 
G_F M_W^2}\cdot T1(\xi_j),\label{czcc}\\ 
T1(\xi_j)&=& \left[ \frac{ \xi_j (-1 -3\xi_j +2\ssa(1+ \xi_j))}{8(1-\xi_j)} 
- \frac{\xi_j^2 \cca }{2(1-\xi_j)^2 } \ln \xi_j  \right]\label{t1xij}\\ 
\eeq
where $\lambda_j=V_{js}^*V_{jd}$, $\xi_t=m_t^{*2}/\mpcc^2$ with 
$m_t^* =(1-\epsilon) m_t$, $\xi_j=m_j^2/\mpcc^2$ for $j=c, u$. 

For the case of unit-charged b-pion $\pdd$, we have 
\beq
\Gamma^{II}_{Z_{\mu}} = 
\frac{1}{16 \pi^2}\frac{g^3}{\cos\theta_W}\; \sum_{j} \lambda_j\,
\overline{s_R}\, \gamma_{\mu}\, d_R\; C_0(\eta_j), \label{bszd}  
\eeq
with 
\beq
C_0(\eta_j) = \frac{a_R^{ts*}a_R^{td}\; \mpdd^2 }{2\sqrt{2}\fpit^2 
G_F M_W^2} \left[ \frac{ \eta_j (-1 +\eta_j +2\ssa(1+ \eta_j))}{8(1-\eta_j)} 
+ \frac{ \eta_j^2 \ssa }{2(1-\eta_j)^2 } \ln[\eta_j] \right]
\label{czdd}
\eeq
where $\eta_t=m_t^{*2}/\mpdd^2$, $\eta_j=m_j^2/\mpdd^2$ for $j=c, u$. 

For the case of technipions $\paa$ and $\pbb$ , we have 
\beq
\Gamma^{III}_{Z_{\mu}} &=& 
\frac{1}{16 \pi^2}\frac{g^3}{\cos\theta_W}\; \sum_{j} \lambda_j\,
\overline{s_L}\, \gamma_{\mu}\, d_L\; C_0(y_j), \label{bsza}  \\
\Gamma^{IV}_{Z_{\mu}} &=& 
\frac{1}{16 \pi^2}\frac{g^3}{\cos\theta_W}\; \sum_{j} \lambda_j\,
\overline{s_L}\, \gamma_{\mu}\, d_L\; C_0(z_j), \label{bszb}  
\eeq
with
\beq
C_0(y_j) &=&\frac{\mpaa^2 }{3\sqrt{2}F_{\pi}^2 G_F M_W^2} 
\cdot T1(y_j), \label{czaa}\\ 
C_0(z_j) &=&\frac{8 \mpbb^2 }{3\sqrt{2}F_{\pi}^2 G_F M_W^2} 
\cdot T1(z_j), \label{czbb}
\eeq
where $y_t=m_{t1}^2/\mpaa^2$ and $z_t=m_{t1}^2/\mpbb^2$ with 
$m_{t1}=\epsilon m_t$, 
and $y_j=m_j^2/\mpaa^2$ and $z_j=m_j^2/\mpbb^2$ for $j=u,c$. 

The  new $C_0$ functions in (\ref{czcc},\ref{czdd},\ref{czaa},\ref{czbb}) 
are  just the same kind of functions as the $\Gamma_Z$ in (2.7) of 
ref.\cite{inami81} or the $C_0(x_i)$ in (2.18) of the first paper in 
ref.\cite{buras974}. Each new $C_0$ function describes the contribution to 
the $\dsz$ vertex from the corresponding scalar.
In the numerical calculations we will include the new contributions 
to the rare K-decays by simply adding the new $C_0$ functions  
with their standard model counterpart $C_0(x_i)$.

In the above calculations, we used the unitary relation $\sum_{j=u,c,t}
\lambda_j\cdot \;constant=0$ wherever possible, and  neglected the masses 
for all external quark lines. We also used the functions $(B_0, B_{\mu}, 
C_0, C_{\mu}, C_{\mu\nu})$ whenever needed to make the integrations, and the
explicit forms of these complicated functions can be found, for instance, 
in the Appendix-A of ref.\cite{cho91}. 

When the new contributions from charged scalars are included, the 
functions $X(x_t)$, $Y(x_t)$, $P_0(X)$ and $P_0(Y)$ appeared in  
eqs.(\ref{bra},\ref{brb},\ref{brcsd}) should be modified as the following,  
\beq
X_{tot} &=& X(x_t) + X^{New}, \label{xxtt} \\
Y_{tot} &=& Y(x_t) + X^{New}, \label{yxtt}\\
P_0(X)_{tot} &=& P_0(X) + P_0^{New}, \label{p0xt} \\
P_0(Y)_{tot} &=& P_0(Y) + P_0^{New}, \label{p0yt} 
\eeq
with
\beq
X^{New} &=& C_0(\xi_t)+ C_0(\eta_t)+ C_0(y_t)+ C_0(z_t), \label{x0new}\\
P_0^{New} &=& \frac{1}{\lambda^4} \left[ C_{NL}(\pcc)+ C_{NL}(\pdd)+ 
C_{NL}(\paa)+C_{NL}(\pbb) \right ], \label{p0new}
\eeq
where the function $X^{New}$ describes  the correction from the dominant 
top quark part, while the function $P_0^{New}$ corresponds to the charm part. 
The new charm part is numerically very small: no more than $2\%$ of the total 
new contribution. 
For the convenience of the reader, we present the explicit expressions of 
functions $C_{NL}(\pi_i)$ ($\pi_i= \pcc, \pdd, \paa, \pbb$) in Appendix B.

In the SM, by using the SIP we have 
\beq
X(x_t)&=&1.537, \ \ Y(x_t)=1.032,\nonumber\\  
P_0(X)&=&0.412, \ \  P_0(Y)=0.155.\label{xysm}
\eeq

For the Case-A, by using the SIP (\ref{sip}) and assuming 
$\mpcc=100GeV$ and $\mpdd=300 GeV$, we have 
\beq
X(\xi_t) &=& 2.258, \ \ P_0^{New}(\pcc) = -0.0146, \\ 
X(\eta_t) &=& -7 \times 10^{-4}, \ \ |P_0^{New}(\pdd)| \leq 10^{-6}, 
\eeq
here it is easy to see that $X(\eta_t)$ and $P_0^{New}(\pdd)$ are clearly 
much smaller than $X(\xi_t)$ and $P_0^{New}(\pcc)$, we therefore neglect 
the contribution from the b-pions $\pdd$ in Case-A. 

For the Case-B, by using the SIP (\ref{sip}) and assuming 
$\mpcc=100GeV$ and $\mpdd=300 GeV$, we have 
\beq
X(\xi_t) &=& 2.258, \ \ P_0^{New}(\pcc) = -0.0146, \\ 
X(\eta_t)&=& \pm 1.746, \ \ P_0^{New}(\pdd)= \pm 0.146, 
\eeq
where the sign of $X(\eta_t)$  and $P_0^{New}(\pdd)$ will be determined by
the sign of $a_R^{ts*}a_R^{td}$. Both $X(\eta_t)$  and $P_0^{New}(\pdd)$ will
be positive (negative) when the product $a_R^{ts*}a_R^{td}$ is negative 
(positive). In the following, we use the term 
Case-B1 and Case-B2 to denote the case of assuming $a_R^{ts*}a_R^{td}= 
1/4, -1/4$, respectively. For the Case-B1 (Case-B2), 
the new contribution from $\pcc$ and $\pdd$ will cancel (enhance) each 
other. 

In TC2 models, the new contribution to the rare 
K-decays from ordinary technipions is strongly suppressed by a factor of 
$(\epsilon \fpit/F_\pi)^2 \sim 10^{-3}$ for $F_\pi \approx 123 GeV$
and $\epsilon \approx 0.05 $, when compared with that from the top-pions. 
Numerically, less than $5\%$ of the total new contribution is due to $\paa$
and $\pbb$. We therefore use the fixed values of 
$\mpaa=50 GeV$ and $\mpbb=100GeV$ in the following numerical calculations. 
For heavier technipions, their contributions will become even smaller. 

In the multiscale walking technicolor model\cite{epj981}, on the contrary, 
the enhancements to the rare K-decays 
due to $\paa$ and $\pbb$ can be as large as $2 \sim 3$ orders of  
magnitudes. The major reason is the big difference in how to generate large 
top quark mass in different models, which in turn result in very different 
effective Yukawa couplings. In the TC2 models, $\paa$ and $\pbb$ couple to 
the top quark with strength $\epsilon m_t/F_\pi \approx 0.1$,  
which is much smaller than the coupling in the MWTCM: $ m_t/F_Q\approx 
4$ for $F_Q \approx 40 GeV$\cite{epj981}. 
And finally the  technipions $\paa$ and $\pbb$ contribute to the rare 
K-decays very differently in the MWTCM and the TC2 models.

Fig.2a and Fig.2b show the $X$ functions for the Case-A and Case-B, 
respectively. The short-dashed line is the standard model prediction 
$X(x_t)=1.537$. In Fig.2a, the long-dashed curve shows the 
function $X(\xi_t)$ and the solid curve corresponds to the function 
$X_{tot}$.  The positive $X(\xi_t)$ greatly enhances the $X(x_t)$ for light 
$\pcc$. In Fig.2b, the dotted and solid curve show the function $X_{tot}$ 
function for Case-B1 and Case-B2, respectively.

Within the range of $\epsilon=0.03 \sim 0.1$, the $X$ functions basically 
remain unchanged. For $\fpit=60 GeV$, the functions $X(\xi_t)$ and 
$X(\eta_t)$ will be decreased by a factor of $(5/6)^2$.

In the following two sections we will present the numerical results for the 
branching ratios $\brka$, $\brkb$ and $\brkcsd$ with the inclusion of new 
physics effects, 
compare the theoretical predictions with the data available currently. 

%%%%%%%%%%%%%
\section{Rare decays $\ka$ and $\kb$}

\subsection{The decay $\ka$}

The rare decay $\ka$ is theoretically very clean and the long-distance 
contributions were known to be negligible\cite{buras974}. 
When the new contributions from scalars are included, one finds
\beq
B(\ka) = \kappa_+ \cdot \left[ \left( \frac{Im \lambda_t}{\lambda^5}
X_{tot} \right)^2 + \left( \frac{Re \lambda_c}{\lambda}P_0(X)_{tot} + 
\frac{Re \lambda_t}{\lambda^5} X_{tot} ) \right)^2 \right] 
\label{brkatot}
\eeq
where $\kappa_+ =4.11 \times 10^{-11}$\cite{buras974}, the functions 
$X_{tot}$ and  $P_0(X)_{tot}$ are given in (\ref{xxtt},\ref{p0xt}).

Using the SIP (\ref{sip}),  and assuming  
$ \mpcc= (100 - 500) GeV$ and $\mpdd= 300GeV$, we have  
\beq
\brka = \left \{ \begin{array}{ll} 
9.39\times 10^{-11} &  {\rm in \ \  SM} \\
(3.92-0.89)\times 10^{-10} & {\rm in \ \  Case-A} \\
(1.95 - 0.13)\times 10^{-10} & {\rm in  \ \ Case-B1}\\ 
(6.59 - 2.34)\times 10^{-10} & {\rm in  \ \ Case-B2} 
\end{array} \right.
\label{brkatp}
\eeq

On the experimental side, the first event for $\ka$ has been recently 
observed by the BNL787 collaboration \cite{adler97}, giving
\beq
\brka_{exp} = 4.2^{+9.7}_{-3.5} \times 10^{-10}, \label{brkaexp}
\eeq
in the ball park of the SM expectations. Further data already collected are 
expected to increase the sensitivity by more than a factor $2$, and there ae 
plans to collect data  representing a further large increase in sensitivity.

Fig.3a and Fig.3b show the $\mpcc$ dependence of $\brka$  in Case-A and 
Case-B, respectively.  In Fig.3, the horizontal 
band between two solid lines corresponds to the  data (\ref{brkaexp}), 
while the short-dashed line is the standard model prediction.
The solid curve in Fig.3a shows the branching ratio $\brka$ when the new 
contributions are included. In Fig.3b, the lower (upper) solid and 
short-dashed curve are the branching ratios $\brka$ in Case-B1 (Case-B2) 
for $\mpdd=300,1000 GeV$, respectively.

From Fig.3b, an upper bound on $\mpcc$ can be read out: 
$\mpcc \leq 285 GeV$ for $\mpdd \leq 1000 GeV$ in Case-B1. This upper 
bound will be weakened by about $50GeV$ if we consider uncertainties of 
other parameters.

From Fig.3 one can see that the theoretical predictions for the branching 
ratio $\brka$ in TC2-I model are now in good agreement with the data
(\ref{brkaexp}) for all three cases.  The uncertainty of the data is
still large. Further improvement of the data will be 
very helpful to constraint the TC2 models from this decay mode.

%%%%%%%%%%%%%%%  
\subsection{  The decay $\kb$ }

In the SM, the rare decay $\kb$ is completely dominated by short-distance 
loop effects with the top quark exchanges and  there is no theoretical 
uncertainties due to $m_c, \mu_c$ and $\Lambda_{\overline{MS}}$ present 
in the decay $\ka$. Consequently this decay mode is even cleaner than $\ka$ 
and is very well suited for the probe of new physics if the experimental 
data can reach the required sensitivity. 

When the new contributions from scalars are included, the branching ratio 
will be 
\beq
\brkb = \kappa_L \cdot \left( \frac{Im \lambda_t}{\lambda^5} X_{tot} 
\right)^2. 
\eeq
where $\kappa_L =1.80\times 10^{-10}$\cite{buras974},  
function $X_{tot}$ is given in (\ref{xxtt}).

Using the SIP (\ref{sip}) and assuming $ \mpcc= 
(100 - 500)GeV$ and $\mpdd= 300 GeV$, we have  
\beq
\brkb = \left \{ \begin{array}{ll} 
2.74\times 10^{-11} &\ \  {\rm in \ \  SM} \\
(1.67 - 0.28)\times 10^{-10} & \ \ {\rm for \ \  Case-A} \\
(0.75 - 0.01) \times 10^{-10} & \ \ {\rm for  \ \ Case-B1} \\
(2.95 - 0.91) \times 10^{-10} & \ \ {\rm for  \ \ Case-B2} 
\end{array} \right.
\eeq

On the experimental side, the KTeV group has recently quoted a preliminary 
result\cite{david97}
\beq
 \brkb \leq 1.8 \times 10^{-6}, \ \ 90\% C.L, \label{brkbexp}
\eeq
and the same group aims at reaching in 
1999 a single event sensitivity of $3\times 10^{-9}$\cite{chen97}. The CLOE 
experiment in $DA\Phi NE$ can also reach the sensitivity of $10^{-9}$ in the 
next few years\cite{bossi99} 

Fig.4 shows the $\mpcc$ dependence of $\brkb$  in all three cases.  
The short-dashed line is the standard model prediction. The middle 
solid curve is the branching ratio in Case-A; while the lower (upper) 
two curves are the branching ratios in Case-B1 (Case-B2) for $\mpdd=300, 
1000GeV$, respectively. The enhancement can be as large as one order of 
magnitude for Case-B2.

Although the current bound (\ref{brkbexp}) is 
still about four orders of magnitudes above the theoretical 
expectation after including the contributions from new scalars, 
it is possible to measure this gold-plated decay mode with enough 
sensitivity to probe the effects of new physics in next few years. 
Sensitivities around $ 10^{-11}$ are the goal of three dedicated 
experiments which have been recently proposed\cite{chen97,bnl96,kek96}. 
A recent proposal\cite{bnl96}, for example, aims to make a $\sim 15\%$ 
measurement of $\brkb$.

%%%%%%%%%%%%%%%%%
\section{ The decay $\kc$ }

The rare decay $\kc$ is a potentially important channel to study the
weak interaction within the SM, as well as possible effects of new physics.
This decay proceeds through two different mechanisms: a dominant 
long-distance (LD) part from the two-photon intermediate state and a 
short-distance (SD) part, which in the SM arises from one-loop 
$Z^0$-penguin and 
W-box diagrams involving  gauge bosons. Since the short-distance part is 
sensitive to the presence of virtual top quark and other new heavy particles
predicted by many new physics models\footnote{In TC2 models, for example, 
the virtual unit-charged scalars will appear in the $Z^0$-penguin diagrams 
as shown in Fig.1, and thus provide new contributions to the rare 
K-decays.}, it offers a window into new physics phenomena. 

For the decay $\kc$, the full branching ratio can be written generally as 
follows\cite{pich98}
\beq
\brkc &=& 2 \beta\; B(K_L \to \gamma \gamma)\left ( 
\frac{\alpha m_{\mu}}{\pi M_K}\right )^2
\left( Re[A]^2 + Im [A]^2 \right) \label{brkcf}\\
Re[A] &=&A_{SD} + A_{LD} 
\eeq
where  $\beta=\sqrt{ 1-4 m_{\mu}^2/M_K^2 }$,  $Im[A]$ denote the absorptive 
contribution arising from a two-photon intermediate state, and finally  
$A_{SD}$ and $A_{LD}$ represent the short- and long-distance dispersive 
contribution, respectively.

In the SM, the short-distance part of $\brkc$ is \cite{buras974} 
\beq
\brkcsd = ( 1.23 \pm 0.57) \times 10^{-9}, \label{brkcsd} 
\eeq
where the error is dominated by the uncertainty of $|V_{cb}|$.  Using the 
measured branching ratio of $B(K_L \to \gamma \gamma)=(5.92 \pm 0.12) 
\times 10^{-4}$\cite{pdg98}, one gets 
\beq
B(K_L \to \mu^+ \mu^-)_{abs} = (7.07 \pm 0.18) \times 10^{-9},
 \label{brkcabs}
\eeq
which is very close to the measured rate\cite{kek137,pdg98}
\beq
\brkc = (7.2 \pm 0.5) \times 10^{-9}. \label{brkcexp}
\eeq
It is easy to see that the rate $\brkc$ is almost 
saturated by the absorptive contribution, leaving only a small room 
for the coherent sum of the long- and short-distance dispersive contribution, 
\beq
B(K_L \to \mu^+ \mu^-)_{dis}^{(LD + SD)} =
(0.1 \pm 0.5) \times 10^{-9}. \label{brkcdis}
\eeq
Therefore, the magnitude of the total real part $Re[A]$ must be 
relatively small compared with the absorptive part \footnote{The data 
constrain only the size of $Re[A]$, and thus leaves an 
ambiguity for the sign of $Re[A]$.}. Such a small total dispersive 
amplitude can be realized either when the $A_{SD}$  and $A_{LD}$ parts 
are both small or by partial cancellation between these two parts.

In ref.\cite{pich98}, the authors estimated the dispersive two-photon 
contribution to the decays $P \to l^+ l^-$ ($P=\pi^0, \eta, K_L$, and 
$l=e, \mu$ ) in the framework of the chiral perturbation theory and 
large-$N_C$ considerations and found that
\beq
Re[ A(P \to l^+ l^-)] &=& \frac{1}{4\beta} \ln^2[\frac{1-\beta}{1+\beta}]
+ \frac{1}{\beta}Li_2[\frac{\beta-1}{\beta+1}]
+ \frac{\pi^2}{12 \beta} + 3\ln[\frac{m_l}{M_\rho}] + \chi(M_\rho)
\nonumber\\
&=&
 \left \{ \begin{array}{ll} 
3.2^{+0.8}_{-1.0},  &  {\rm for } \ \  \eta \to \mu^+ \mu^- \\
2.9^{+0.8}_{-1.0} - A_{SD}, &  {\rm for } \ \ K_L \to \mu^+ \mu^- 
\end{array} \right.
\label{rea}
\eeq
where  $M_\rho=0.77GeV$ is the mass of $\rho$ meson, $\chi(M_\rho)=
5.5^{+0.8}_{-1.0}$ is the local contribution  determined by fitting 
the measured ratio $B(\eta \to \mu^+ \mu^-) =(5.8 \pm 0.8)\times 10^{-6}$
\cite{pdg98}. The relative sign between the short- and long-distance 
dispersive amplitude in eq.(\ref{rea}) is fixed by the known positive 
sign of $g_8$ in the large-$N_c$ limit\cite{pich96}. 
From the measured branching ratios of $\eta \to \mu^+ \mu^-$ and $K_L \to 
\mu^+ \mu^-$ decays,  constraints on the short-distance part $A_{SD}$ of 
the decay $\kc$ were derived \cite{pich98}:
\beq
A_{SD} = \left \{ \begin{array}{ll} 
2.2 ^{+ 1.1}_{-1.3}, &  {\rm for }\ \  Re[A] > 0 \\
3.6 \pm 1.2, &  {\rm for }\ \  Re[A] < 0 
\end{array} \right. \label{asdexp}
\eeq
The first bound is in good agreement with the standard model expectation 
$A_{SD}^{SM}=1.8 \pm 0.6$\cite{pich98}, while the second bound shows a 
discrepancy of about $1.4 \sigma$ with the SM expectation. However, the 
errors of above two bounds could be reduced by improving the measurements of 
the branching ratio $\eta \to \mu^+ \mu^-$ and $K_L \to \mu^+ \mu^-$. 
From eqs.(\ref{rea}) and (\ref{asdexp}), one can see that 
the short- and long-distance dispersive contributions are comparable in size 
but  cancel each other strongly. Above bounds on $A_{SD}$ can be  
translated into the constraints on $\brkcsd$ directly, 
\beq
\brkcsd = \left \{ \begin{array}{ll} 
(0.2 - 2.8) \times 10^{-9}, &  {\rm ( Bound-1:\ \ for \ \ } Re[A] > 0 ) \\
(1.5 - 6.0) \times 10^{-9}, &  {\rm ( Bound-2:\ \ for \ \ } Re[A] < 0 )  
\end{array} \right. \label{brkcsdt}
\eeq
Obviously, the bounds are still relatively weak because of the sign 
ambiguity of $Re[A]$. 

In the SM, the branching ratio $\brkcsd$ is known at the NLO 
level\cite{buras974}, as given in (\ref{brkcsd}).  By comparing 
the above bounds on $\brkcsd$ with the theoretical predictions 
after including new physics contributions, one may find useful information 
on TC2 models.

When the new contributions from scalars in TC2-I model are included, we have
\beq
\brkcsd = \kappa_{\mu} \cdot \left[ \frac{Re \lambda_c}{\lambda}P_0(Y)_{tot} 
+ \frac{Re \lambda_t}{\lambda^5}Y_{tot} \right]^2.
\eeq
where $\kappa_{\mu} =1.68\times 10^{-9}$\cite{buras974}, the functions 
$Y_{tot}$ and  $P_0(Y)_{tot}$ are given in (\ref{yxtt},\ref{p0yt}).
Using the SIP (\ref{sip}) and assuming $ \mpcc= (100- 500)GeV$ 
and $\mpdd= 600 GeV$, one gets  
\beq
\brkcsd = \left \{ \begin{array}{ll} 
1.25\times 10^{-9} &  {\rm in \ \  SM} \\
(9.18 - 1.13) \times 10^{-9} & {\rm for \ \  Case-A} \\
(4.57 - 0.03) \times 10^{-9} & {\rm for  \ \ Case-B1} \\
(15.36 - 3.81)\times 10^{-9} & {\rm for  \ \ Case-B2} 
\end{array} \right. \label{brkcsdth}
\eeq
The enhancement can reach one order of magnitude. 

Fig.5a shows the branching ratio $\brkcsd$ versus the mass $\mpcc$ for the 
Case-A and Case-B1. The short-dashed line is the SM prediction,  while 
the horizontal band here corresponds to the Bound-1.
The upper solid curve in Fig.5a shows the branching ratio $\brkcsd$ in 
Case-A, and a lower bound $\mpcc \geq 280 GeV$ can be read out.  
The lower three curves in Fig.5a are the ratios   
$\brkcsd$ in Case-B1 for $\mpdd=300, 600, 1000GeV$ respectively.

Fig.5b shows the same thing as the Fig.5a, but the horizontal band here 
stands for the Bound-2. For the Case-A, the theoretical 
prediction is consistent with the bound, while the bound-2 prefer light 
top-pions for Case-B1: $\mpcc \leq 230 GeV$ for $\mpdd \leq 1000 GeV$.  

Fig.6a and Fig.6b show the same thing as  Fig.5a and Fig.5b but for the 
Case-B2 instead. The upper three curves in Figs.(6a) and (6b) are the ratios   
$\brkcsd$ in Case-B2 for $\mpdd=300, 600, 1000GeV$ respectively.
The Case-B2 is completely excluded if $Re[A] > 0$, but is still allowed  
if $Re[A] < 0 $ and $\mpcc \geq 270 GeV$. 

For the decay $\kc$, there is a strong cancellation between the short- and 
long-distance dispersive parts. For the Case-A, the enhancement to $\brkcsd$
can reaches a factor of 8 for light top-pions. For the Case-B1 (Case-B2), 
the new contributions from top-pions and b-pions will cancel (enhance) each 
other, and the resultant enhancement can reaches a factor of 4 (12) for light 
top-pions.

%%%%%%%%%%%%%%%%%  
\section{ Conclusion and discussions}

In this paper we calculated the new contributions to the rare FCNC K-decays 
$\ka$, $\kb$ and $\kc$ from the new  $Z^0$-penguin and box diagrams  induced 
by the unit-charged top-pions $\pcc$, b-pions $\pdd$,  and technipions 
$\paa$ and $\pbb$ appeared in the TC2 models. We choose the TC2-I model 
proposed by Hill \cite{hill95} as the typical TC2 model to do the analytical 
and numerical calculations. It is beyond the scope of this paper to consider 
the detailed differences between TC2 models.

From the analytical evaluations of the one-loop Feynman diagrams, we extract 
out the new functions $C_0(\pi_i)$ and $C_{NL}(\pi_i)$ ($\pi_i = \pcc, \pdd, 
\paa, \pbb$) which describe the new $Z^0$-penguin contributions due to 
unit-charged scalars, combine the new functions with their standard 
model counterparts and use them directly in the calculation of branching 
ratios. The $m_t$-dependent term $X^{New}$ in (\ref{x0new}) dominates over 
the $m_t$-independent term $P_0^{New}$ in (\ref{p0new}).

The mixing matrixes $D_L$ and $D_R$ also play an important role for the 
characteristics  and magnitudes of the new contributions. But unfortunately, 
these new mixing matrices are really the most undetermined part of the TC2 
models. Thanks to the accurate  experimental measurement of 
$B^0 - \overline{B^0}$ mixing, we got some strong constraints 
(\ref{arl1},\ref{arl3}) on the relevant mixing factors. The Case-A is 
allowed by the constraints (\ref{arl1},\ref{arl3}), while the Case-B1 and 
Case-B2 are also considered for the purpose of  comparison and illustration.

For the decay $\ka$, the enhancement to the ratio $\brka$ can reach a factor 
of $2 \sim 7$. The theoretical predictions in TC2-I model are generally agree 
well with the data (\ref{brkaexp}) for all three cases. Of course, 
the uncertainty of the data is still very large and further improvement of 
the data will be very helpful to test or constraint the TC2 models from this 
decay mode.

The decay $\kb$ is the cleanest decay mode among  three decay modes in 
question. The enhancement to the branching ratio $\brkb$ due to the 
top-pions and b-pion can be as large as one order of magnitude. 
But the central 
problem for this decay mode is the very low sensitivity of the available 
data, which is about four orders of magnitudes above the theoretical 
expectation. Further improvements of the data will be very essential to find 
the signals of new physics through this decay mode. 

For the decay $\kc$, the situation becomes more complicated because of the 
involvement of the long-distance contributions.
After including the additional short-distance part from new 
physics, the theoretical predictions are still consistent with 
the data for Case-A and Case-B1. The Case-B2, however, is disfavored by the 
data. The major obstacles in extracting strong constraints on $A_{SD}$ out of 
the decay $\kc$ is the large uncertainty of $A_{LD}$, the sign ambiguity of 
$Re[A]$ as well as the strong cancellation between the short- and 
long-distance dispersive parts. And therefore  
improvements in theoretical predictions and the experimental data will be 
very essential for us to test the new physics effects through this decay 
mode.

In summary, the unit-charged scalars appeared in TC2 models can provide  
sizable new contributions to the rare K-decays $\ka$, $\kb$ and $\kc$ through 
$Z^0$-penguin diagrams. The accurate  data of $B^0 - \overline{B^0}$ 
mixing lead to strong constraints on the size of $D_R^{bd}$ and $D_R^{bs}$
if we use the square-root ansatz for the $D_L$.  Some simple but interesting 
lower or  upper bounds on $\mpcc$ are obtained by comparing the theoretical 
predictions with the relevant data. The TC2-I model is,  in general,  still 
consistent with the available data of rare K-decays in 
question. Further improvement of the data and the theoretical predictions 
will be very helpful to constraint the TC2 models from the rare K-decays.

\section*{ACKNOWLEDGMENT}
This work is  supported in part by the National Natural Science Foundation of 
China under the Grant No.19575015, and by the Outstanding Young Teacher 
Foundation of the Education Ministry of China, as well as the funds from 
Henan Science and Technology Committee.

\section*{Appendix A}

In this Appendix, we present the explicit expressions for the functions 
$B_0(x_t)$, $C_0(x_t)$, $X_1(x_t)$, $Y_1(x_t)$. 
The functions of $C_0(x_t)$ and $B_0(x_t)$ govern the leading top 
quark contributions through the $Z^0$-penguin and W-box diagrams 
in the SM, while the functions $X_1(x_t)$ and $Y_1(x_t)$ describe
the NLO QCD corrections,
$$
B_0(x_t)= \frac{1}{4} \left[ \frac{x_t}{1-x_t} 
+ \frac{x_t \ln[x_t]}{(x_t-1)^2} \right] \hspace{4cm} \eqno{(A1)}
$$$$
C_0(x_t)= \frac{x_t}{8} \left[ \frac{x_t-6}{x_t-1} 
+ \frac{3x_t + 2}{(x_t-1)^2}\, \ln[x_t] \right] \hspace{4cm} \eqno{(A2)}
$$
$$
X_1(x_t) = -\frac{23x_t + 5x_t^2 -4x_t^3}{3(1-x_t)^2} 
+ \frac{x_t -11x_t^2 +x_t^3 + x_t^4}{(1-x_t)^3}\ln[x_t] \hspace{3cm} 
$$$$
\hspace{2cm} +  \frac{8x_t +4x_t^2+x_t^3 - x_t^4}{2(1-x_t)^3}\ln^2[x_t]
- \frac{4x_t -x_t^3}{(1-x_t)^2}L_2(1-x_t) 
+ 8x_t\frac{\partial X_0(x_t)}{\partial x_t}\ln[x_{\mu}] \eqno{(A3)}
$$$$
Y_1(x_t) = -\frac{4x_t + 16x_t^2 +4x_t^3}{3(1-x_t)^2} 
- \frac{4x_t -10x_t^2 -x_t^3 - x_t^4}{(1-x_t)^3}\ln[x_t]\hspace{3.4cm}   
$$$$
\hspace{2cm} +  \frac{2x_t -4x_t^2+x_t^3 - x_t^4}{2(1-x_t)^3}\ln^2[x_t]
- \frac{2x_t +x_t^3}{(1-x_t)^2}L_2(1-x_t) 
 + 8x_t\frac{\partial Y_0(x_t)}{\partial x_t}\ln[x_{\mu}] \hfill \eqno{(A4)}
$$
where $x_t=m_t^2/m_W^2$,  $x_{\mu}=\mu^2/M_W^2$ with 
$\mu = {\cal O}(m_t)$ and 
$$
L_2(1-x_t)= \int_1^{x_t}\; dy\, \frac{ln[y]}{1-y}. \eqno{(A5)}
$$

For the charm sector in the SM, the $C_{NL}$ is the $Z^0$-penguin part and 
the $B_{NL}^{1/2}$ ($B_{NL}^{-1/2}$) is the box contribution, 
relevant for the case of final state leptons with $T_3=1/2$ ($T_3=-1/2$): 
$$
C_{NL} = \frac{x(m)}{32}\,K_c^{24/25}\,\left[ \left(
\frac{48}{7}K_+ +\frac{24}{11}K_-  -\frac{ 696}{77}K_{k33} 
\right)\left(\frac{4\pi}{\alpha_s(\mu)} + \frac{15212}{1875}( 1-K_c^{-1}) 
\right) \right. 
$$$$
 \hspace{1cm} +  \left( 1-\ln \frac{\mu^2}{m^2} \right) (16K_+ -8K_-) 
  - \frac{1176244}{13125}K_+ -\frac{2302}{6875}K_-  +  
\frac{3529184}{48125}K_{33} 
$$$$
\hspace{1cm} + \left. K\, \left( \frac{56248}{4375}K_+ -\frac{81448}{6875}K_- 
+\frac{4563698}{144375}K_{33} \right) \right] \eqno{(A6)}
$$
where
$$
K = \frac{\alpha_s(M_W)}{\alpha_s(\mu)}, \ \ 
K_c = \frac{\alpha_s(\mu)}{\alpha_s(m)}, \ \ 
K_+ = K^{6/25},\ \  K_- = K^{-12/25},\ \  K_{33} = K^{-1/25} \eqno{(A7)}
$$
and 
$$
B_{NL}^{1/2}= \frac{x(m)}{4}\,K_c^{24/25}\,\left[3(1-K_2)\left( 
\frac{4\pi}{\alpha_s(\mu)}
   + \frac{15212}{1875}(1-K_c^{-1}) \right) \right.
$$$$
\hspace{1cm} -  \left. \ln\frac{\mu^2}{m^2}
   - \frac{r\ln r}{1-r} -\frac{305}{12} + \frac{15212}{625}K_2 
   + \frac{15581}{7500} K\, K_2 \right] 
\eqno{(A8)}
$$$$
B_{NL}^{-1/2}= \frac{x(m)}{4}\,K_c^{24/25}\,\left[
3(1-K_2)\left( \frac{4\pi}{\alpha_s(\mu)}
   + \frac{15212}{1875}(1-K_c^{-1}) \right) \right. 
$$$$
\hspace{1cm}  -  \left. 
\ln\frac{\mu^2}{m^2}-\frac{329}{12} + \frac{15212}{625}K_2 
   + \frac{30581}{7500} K\, K_2 \right]  \eqno{(A9)}
$$
here $K_2=K_{33}$, $m= m_c$, $\mu = {\cal O}(m_c)$, $x(m)=m_c^2/M_W^2$,  
$r=m_l^2/m_c^2(\mu)$ and $m_l$ is the lepton mass. 

\section*{Appendix B}

For the dominant top sector in the TC2-I model, the $C_0(\pi_i)$ 
($\pi_i = \pcc, \pdd, \paa, \pbb$) functions 
have been given in (\ref{czcc}$-$\ref{czdd}).

For the charm sector in the TC2-I model, the $C_{NL}(\pi_i)$ 
($\pi_i = \pcc, \pdd, \paa, \pbb$) functions take the form
$$
C_{NL}(\pi_i) = a_i \frac{m_c^2}{m_W^2}K_c^{24/25}\left[ \left(
\frac{48}{7}K_+ +\frac{24}{11}K_-  
-\frac{ 696}{77}K_{33} 
\right)\left(\frac{4\pi}{\alpha_s(\mu)} + \frac{15212}{1875}( 1-K_c^{-1}) 
\right) \right. 
$$$$
\hspace{1cm} +  \left( 1-\ln \frac{\mu^2}{m^2} \right) (16K_+ 
-8K_-)   - \frac{1176244}{13125}K_+ -\frac{2302}{6875}K_-  +  
\frac{3529184}{48125}K_{33} 
$$$$
\hspace{1cm} + \left. K_{\pi_i}\, \left( \frac{56248}{4375}K_+ 
-\frac{81448}{6875}K_- 
+\frac{4563698}{144375}K_{33} \right) \right]. \eqno{(B1)}
$$
with 
$$
K_c = \frac{\alpha_s(\mu)}{\alpha_s(m_c)}, K_{\pi_i} = 
\frac{\alpha_s(m_{\pi_i})}{\alpha_s(\mu)}, \ \ 
K_+ = (K_{\pi_i})^{6/25}, 
$$$$
 K_- = (K_{\pi_i})^{-12/25},\ \  
K_{33} = (K_{\pi_i})^{-1/25}, \ \ \mu = {\cal O}(m_c).  \eqno{(B2)}
$$

For top-pion $\pcc$ and b-pion $\pdd$ we have 
$$
a_1 = \frac{a_{L}^{ts*}a_L^{td}}{64\, \sqrt{2}\,\fpit^2\,G_F}, \ \  
K_{\tilde{\pi}} = \frac{\alpha_s(\mpcc)}{\alpha_s(\mu)},   \ \
a_2  = \frac{a_R^{ts*}a_R^{td}}{8 \sqrt{2}\,\fpit^2\,G_F},  \ \  
K_{\tilde{H}} = \frac{\alpha_s(\mpdd)}{\alpha_s(\mu)}.  \eqno{(B3)}
$$

For ordinary technipions $\paa$ and $\pbb$ we have 
$$
a_3 = \frac{1}{96\, \sqrt{2}\,F_{\pi}^2\,G_F},  \ \  
K_{\paa} = \frac{\alpha_s(\mpaa)}{\alpha_s(\mu)}, \ \   
a_4  = \frac{1}{12\, \sqrt{2}\,F_{\pi}^2\,G_F},  \ \  
K_{\pbb} = \frac{\alpha_s(\mpbb)}{\alpha_s(\mu)}. \eqno{(B4)}
$$

\newpage
\begin{center}
{\bf Figure Captions}
\end{center}
\begin{description}

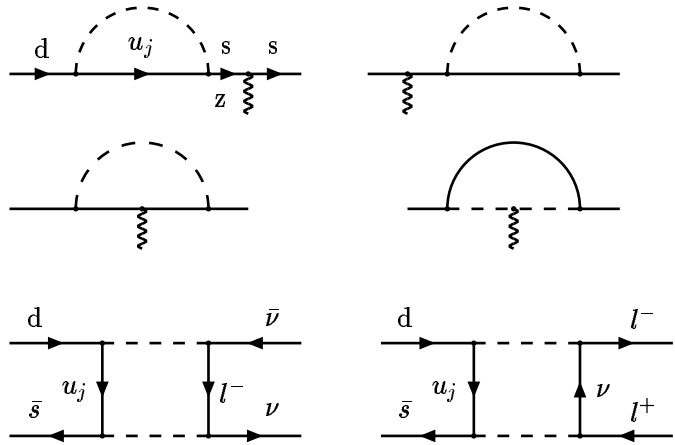
\item[Fig.1:]  The new $Z^0-$penguin and box diagrams induced 
by the internal exchanges of the unit-charged scalars $\paa$, $\pbb$, 
$\pcc$ and $\pdd$. The dashed lines are scalars and the 
$u_j$ stands for the up-type quarks $(u, c, t)$.

\item[Fig.2:]  Plots of the $X$ functions versus the mass $\mpcc$ in TC2-I 
model. For more details see the text.

\item[Fig.3:] The branching ratio $\brka$ in TC2-I model as a 
function of $\mpcc$ for the Case-A and Case-B. The short-dashed line is the 
SM prediction, while the horizontal band shows the data (\ref{brkaexp}). 
The solid curve in (3a) is the branching ratio in Case-A, while the lower 
(upper) two curves in Fig.3b show the ratios in Case-B1 (Case-B2) for 
$\mpdd=300, 1000 GeV$, respectively. 

\item[Fig.4:] The branching ratio $\brkb$ in TC2-I model as functions of 
$\mpcc$. The  short-dashed line is the SM prediction. The long-dashed curve
is the ratio in Case-A, while the lower (upper) two curves show the ratios 
in Case-B1 (Case-B2) for $\mpdd=300, 1000 GeV$, respectively.

\item[Fig.5:] The branching ratio $\brkcsd$ as  a function of $\mpcc$ for  
Case-A and Case-B1. The  short-dashed line is the SM prediction. The 
horizontal band  in Fig.5a (Fig.5b) corresponds to the Bound-1 (Bound-2). 
The upper solid curve stands for the ratio in Case-A, while the lower three 
curves show the ratios in Case-B1 for $\mpdd=300, 600, 1000 GeV$, 
respectively.

\item[Fig.6:] The same as in Fig.5, but for the Case-B2.

\end{description}

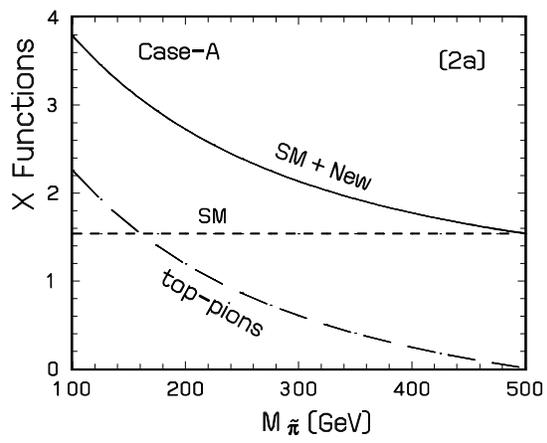
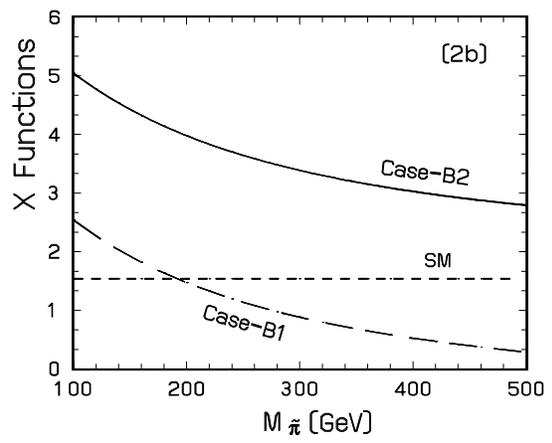
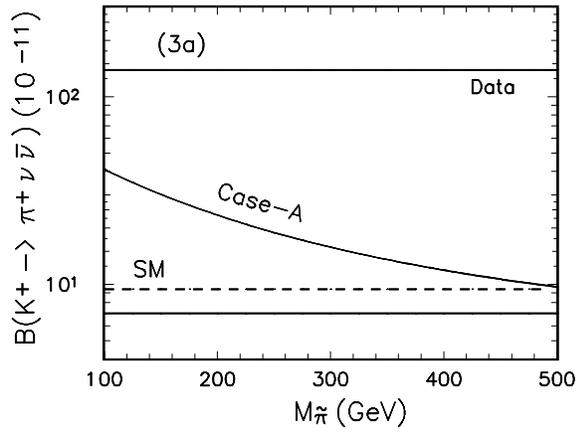
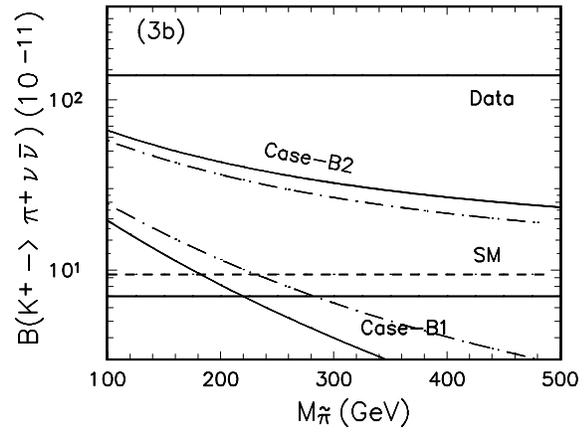
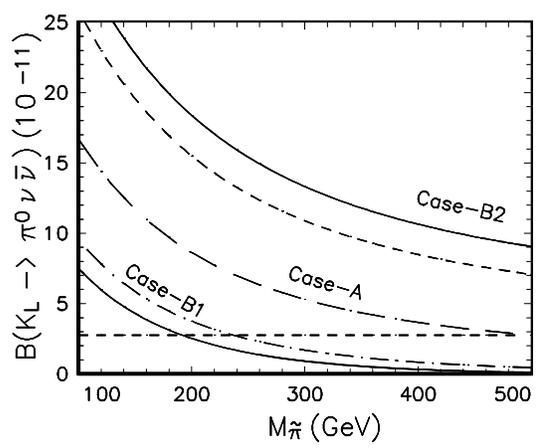
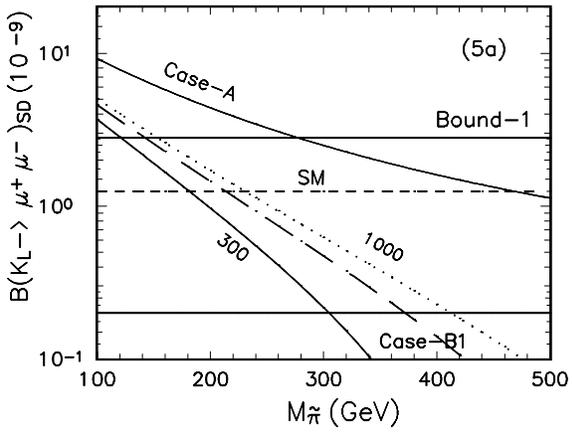
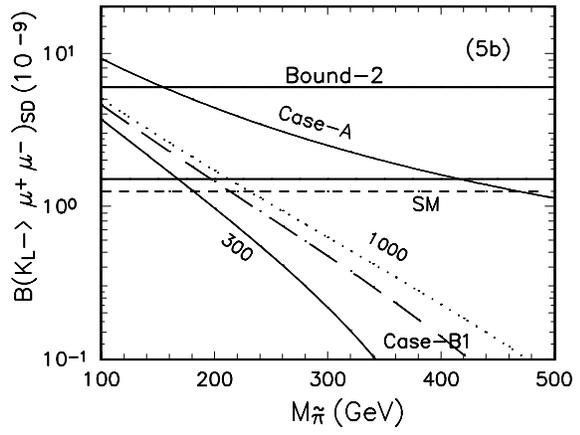
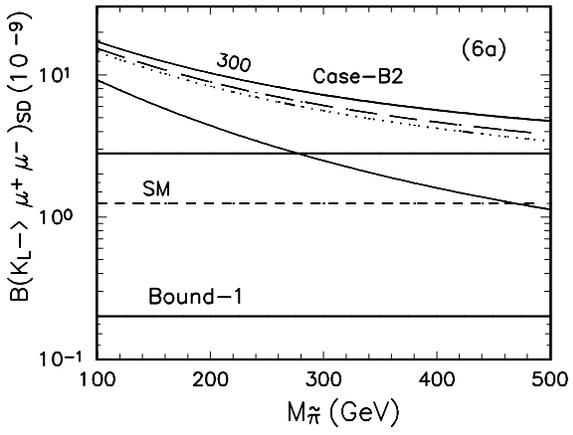
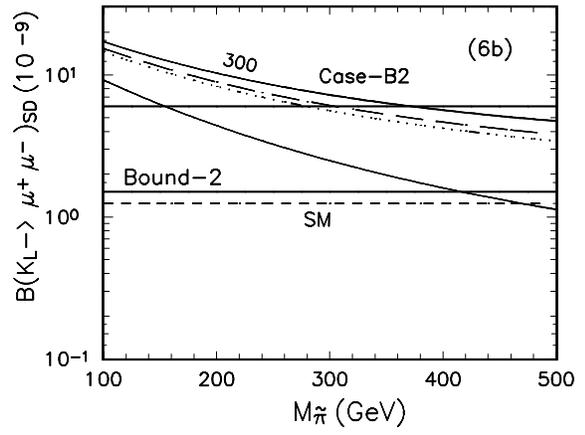
\end{document}